\documentclass[9pt]{extarticle}
\usepackage{spconf,graphicx}
\usepackage{multirow}
\usepackage{booktabs}
\usepackage{ctable}
\usepackage{cite}
\usepackage{amsmath, amssymb, amsthm, enumitem}

\title{Conditional Latent Diffusion-Based Speech Enhancement Via Dual Context Learning}
\name{%
\begin{tabular}{@{}c@{}}
Shengkui Zhao  \quad 
Zexu Pan \quad
Kun Zhou \quad
Yukun Ma\quad 
Chong Zhang  \quad 
Bin Ma
\end{tabular}}
\address{Tongyi Lab, Alibaba Group, Singapore\\
	\{shengkui.zhao, b.ma\}@alibaba-inc.com\\}
\begin{document}
\setlength{\abovedisplayskip}{3pt}
\setlength{\belowdisplayskip}{3pt}
%
\maketitle
\begin{abstract}
Recently, the application of diffusion probabilistic models has advanced speech enhancement through generative approaches. However, existing diffusion-based methods have focused on the generation process in high-dimensional waveform or spectral domains, leading to increased generation complexity and slower inference speeds. Additionally, these methods have primarily modelled clean speech distributions, with limited exploration of noise distributions, thereby constraining the discriminative capability of diffusion models for speech enhancement. To address these issues, we propose a novel approach that integrates a conditional latent diffusion model (cLDM) with dual-context learning (DCL). Our method utilizes a variational autoencoder (VAE) to compress mel-spectrograms into a low-dimensional latent space. We then apply cLDM to transform the latent representations of both clean speech and background noise into Gaussian noise by the DCL process, and a parameterized model is trained to reverse this process, conditioned on noisy latent representations and text embeddings.  By operating in a lower-dimensional space, the latent representations reduce the complexity of the generation process, while the DCL process enhances the model's ability to handle diverse and unseen noise environments.  Our experiments demonstrate the strong performance of the proposed approach compared to existing diffusion-based methods, even with fewer iterative steps, and highlight the superior generalization capability of our models to out-of-domain noise datasets. 
\end{abstract}   
\section{Introduction}
\label{sec:intro}
Speech enhancement, referring to the task of restoring clean speech from noise-corrupted speech signals, has been seen in many applications in telecommunication and robust speech recognition. Recent advancements in speech enhancement have been achieved by leveraging deep learning techniques. There are two typical types of speech enhancement approaches based on the fundamental deep learning models: the discriminative approach and the generative approach. The discriminative approach aims to learn the boundary between clean and noisy speech by directly predicting the clean speech components from the noisy speech \cite{Luo2019, Fu2022M}. 
On the other hand, the generative approach aims to model the underlying distributions of clean speech conditioned on the noisy speech, showing more robust to unseen noise scenarios \cite{Lu2022C, Richter2023S, Lemercier2022}.

Among existing generative models, diffusion probabilistic models \cite{Sohl-Dickstein2015, Ho2020D, Nichol2021} have recently made significant progress in speech enhancement. Diffusion probabilistic models employ a forward process to gradually transform data into a tractable prior, typically a standard Gaussian distribution, and train a neural network to perform the reverse process, generating clean data from this prior. For speech enhancement, these models facilitate the conditional generation of clean speech with the noisy speech as conditioner. In CDiffuSE \cite{Lu2022C}, a generalized conditional diffusion probabilistic model is proposed to utilize a discrete diffusion process for time-domain speech signals. SGMSE+ \cite{Richter2023S} introduces a continuous stochastic differential equation (SDE) based diffusion process in the complex spectrogram domain. However, these models face a heavy computational burden due to the numerous iterative generation steps required for reverse diffusion, further aggravated by the high dimensions of the data space.
To address this issue, StoRM \cite{Lemercier2022} combines dual predictive and generative models to reduce computational demands. The predictive model provides an enhanced version of the noisy speech condition for the diffusion reverse process, aiming to minimize the number of iterative steps. However, the predictive models may introduce additional distortions to the speech generation.

While previous studies focus on estimating clean speech distributions conditioned solely on noisy speech, some approaches, such as NADiffuSE \cite{wang2023N} and NASE \cite{hu2024N}, extract noise-specific information using an auxiliary noise classification network as an additional condition for the diffusion inverse process. It has been demonstrated that the learned features from  classification networks provide additional information that are useful to guide the generation process. However, these classification networks only concentrate on noise types rather than noise distributions, and the necessity for noise labels may also constrain the creation of large-scale training datasets.
 
In this work, we propose a conditional latent diffusion model (cLDM) with a dual-context learning (DCL) framework for speech enhancement. Our cLDM leverages a latent diffusion model with noisy speech conditioning, inspired by recent studies on latent diffusion models (LDMs) for text-to-image synthesis \cite{rombach2022h}, text-to-audio synthesis \cite{liu2023a}, and audio editing \cite{wang2023a}. These studies have demonstrated that diffusion models are highly efficient in lower-dimensional latent spaces and learn more effectively from less redundant, more compact representations.
We train a universal variational autoencoder (VAE) to compress mel-spectrograms into a latent space, where cLDM is then applied for speech generation. To enhance the learning process, we apply DCL to guide the cLDM to model both speech and noise latent distributions, conditioned on the noisy latent space and an encoded text prompt. 
\begin{figure*}[t]
  \centering
  \includegraphics[width=17cm]{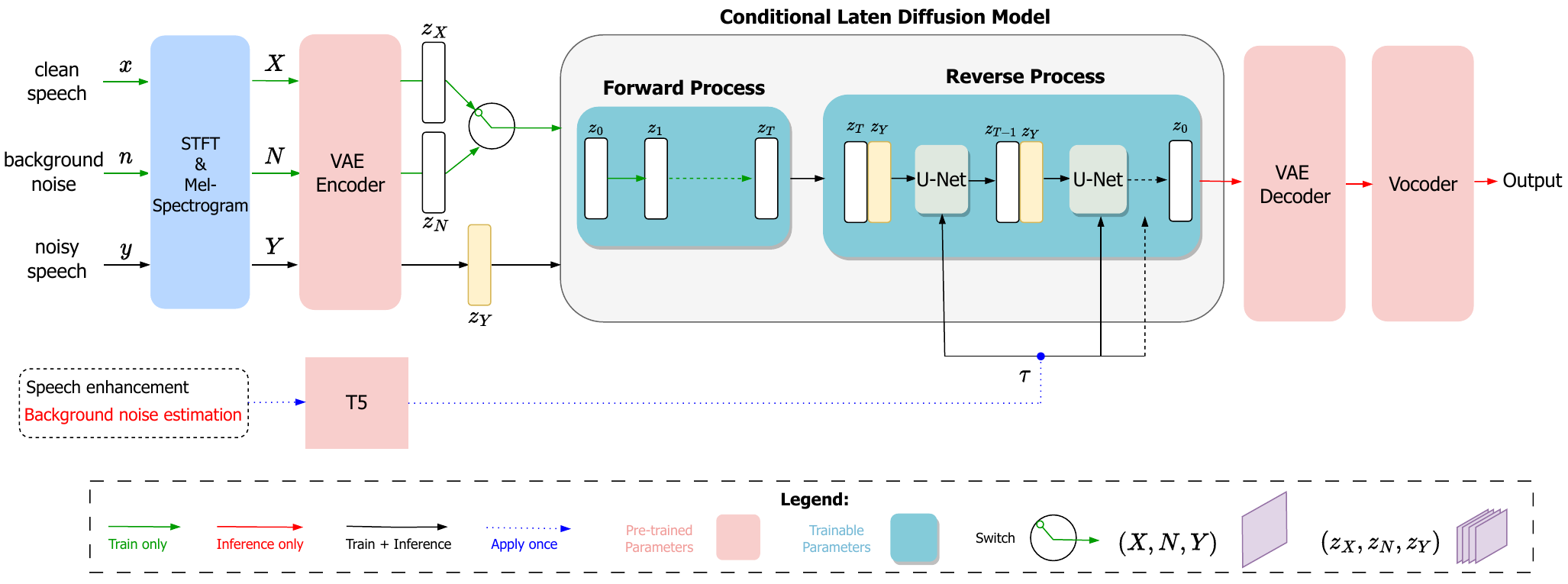}
  \vspace{-3mm}
  \caption{Overview of our proposed cLDM-based speech enhancement system with dual-context learning (DCL) scheme. }
  \label{fig1}
\vspace{-3mm}
\end{figure*} 
By learning the noise distribution, the cLDM becomes aware of the noise characteristics and utilizes this information during the enhancement process. This approach enables the model to accurately identify noise distributions, thereby increasing its discriminative capability and improving the overall clarity and intelligibility of the enhanced speech signal.
Our experiments demonstrate that this approach not only optimizes the generation process but also outperforms other diffusion-based methods, particularly when tested on unseen noises.
\section{Method}
\subsection{System Overview}
The process of speech enhancement involves estimating clean speech $x$ from noisy speech $y: =x+n$, where $n$ is the background noise. The overall architecture of our proposed system is illustrated in Fig. \ref{fig1}. The system primarily consists of a variational auto-encoder (VAE), a conditional latent diffusion model (cLDM), and a vocoder. First, we employ the short-time Fourier Transform (STFT) to convert $x$, $y$ and $n$ into the spectral domain and then extract their mel-spectrograms, denoted by $\{X, Y, N\}\in\mathcal{R}^{L\times F}$ where $L$ is the time dimension and $F$ is the frequency dimension. We then apply the VAE encoder to project the mel-spetrograms $\{X, Y, N\}$ to low-dimensional latent representations  $\{z_X, z_Y, z_N\}\in\mathcal{R}^{C\times \frac{L}{r} \times \frac{F}{r}}$, where $C$ is the channel dimension and $r$ denotes the compression level. During training, the cLDM learns the distributions of $z_X$ and $z_N$, guided by the text embedding $\tau$. For inference, the cLDM generates the corresponding speech prior $z_X$ conditioned on $z_Y$ and $\tau$. Finally, the speech prior $z_X$ is decoded by the VAE decoder and converted back into the waveform domain by the vocoder.  

\subsection{Conditional Latent Diffusion Model}
We use the probabilistic generative model cLDM, as depicted in Fig. \ref{fig1}, to approximate the true conditional data distribution $q(z_0|z_Y,\tau)$ with the learned model distribution $p_\theta(z_0|z_Y,\tau)$, where the initial latent variable 
$z_0$ corresponds to either $z_X$ or $z_N$, depending on the modelled data distribution. To achieve this, cLDM incorporates both forward and reverse processes. 
The forward process diffuses the data distribution into a standard Gaussian distribution in a Markov chain with the following transition probability: 
\begin{align}
q(z_t|z_{t-1})&=\mathcal{N}(z_t;\sqrt{1-\beta_t}z_{t-1}, \beta_t\mathbf{I}),\\
q(z_t|z_0)&=\mathcal{N}(z_t;\sqrt{\bar{\alpha}_t}z_0,(1-\bar{\alpha}_t)\mathbf{\epsilon}),
\end{align}
where $\mathbf{\epsilon} \sim \mathcal{N}(\mathbf{0}, \mathbf{I})$ denotes the injected noise, following a noise schedule $0 < \beta_1 < \beta_2 < \cdots < \beta_T < 1$ with the noise level defined as $\bar{\alpha}_t$ $:=\prod_{i=1}^{t}(1-\beta_i)$ at each step. From the closed form expression of $q(z_t|z_0)$ in Eq. (2), we can directly sample any $z_t$ from the prior $z_0$ via a non-Markovian process:
\begin{equation}
z_t=\sqrt{\bar{\alpha}_t}z_0+(1-\bar{\alpha}_t)\mathbf{\epsilon}.
\end{equation}
As $t$ approaches a sufficient large $T$, $\bar{\alpha}_t$ converges to $0$, and the forward process results in the latent Gaussian distribution $z_T\sim \mathcal{N}(\mathbf{0}, \mathbf{I})$. 

Speech enhancement is achieved by adapting the cLDM for conditional speech generation in the reverse process. The cLDM refines the speech through successive iterations $z_{[0:T-1]}$ based on the learned conditional transition distributions:
\begin{equation}
p_\theta(z_{t-1}|z_t, z_Y, \tau)=\mathcal{N}(z_{t-1};\mathbf{\mu}_\mathbf{\theta}^{(t)}(z_t,z_Y,\tau),\mathbf{\sigma}_t^2\mathbf{I}),
\end{equation}
where the mean and variance are parametrized as follows \cite{Ho2020D}:
\begin{align}
\mathbf{\mu}_\mathbf{\theta}^{(t)}(z_t,z_Y,\tau)&=\frac{1}{\sqrt{\alpha_t}}\left(z_t-\frac{\beta_t}{\sqrt{1-\bar{\alpha}_t}}\hat{\mathbf{\epsilon}}_\mathbf{\theta}^{(t)}(z_t,z_Y,\tau)\right), \\
\mathbf{\sigma}_t^2&=\frac{1-\bar{\alpha}_{t-1}}{1-\bar{\alpha}_t}\beta_t,
\end{align}
where $\hat{\mathbf{\epsilon}}_\mathbf{\theta}^{(t)}(z_t,z_Y,\tau)$ is the parameterized noise estimation using the U-Net model \cite{rombach2022h}, optimized by the following reweighted training loss:
\begin{equation}
\mathcal{L}_{cLDM} = \sum_{t=1}^{T}\gamma_t\mathbb{E}_{\mathbf{\epsilon}_t\sim\mathcal{N}(\mathbf{0},\mathbf{I}),z_0}\|\mathbf{\epsilon}-\hat{\mathbf{\epsilon}}_\mathbf{\theta}^{(t)}(z_t,z_Y,\tau)\|,
\end{equation}
where $\gamma_t$ denotes the weight of reverse step $t$. We incorporate the text guidance $\tau$ into the model using a cross-attention mechanism, and concatenate the latent condition $z_Y$ with $z_t$ at the channel level. As a result, the number of input channels in the first layer of the U-Net is twice the number of output channels in the last layer.
\subsection{Dual-Context Learning Scheme}
Prior works \cite{wang2023N, hu2024N} exploit the noise information through  noise classifiers for noise type classifications, which not only requires the noise labels limiting the scale of training data but also cannot fully take noise variations within each noise type. In this work, we further exploit the noise information in the latent domain to improve the reverse denoising process. In particular, we train a shared cLDM to be capable of generating both the speech prior $z_X$ and the noise prior $z_N$. To achieve this, we use generated noisy-clean data $\mathcal{D}_{y,x}=\{y_i,x_i,instructA\}_{i=1}^{M}$ and noisy-noise data $\mathcal{D}_{y,n}=\{y_i,n_i, instructB\}_{i=1}^{O}$ to train a shared cLDM in a randomly selected manner. 
We select the text of ``{Speech enhancement}" for $instructA$ and ``{Background noise estimation}" for $instructB$ to guide the generation process. The texts are then converted into embeddings using a pre-trained T5 model \cite{Raffel2020E}. Both embeddings are used during training and only the embedding from $instructA$ is used for inference."

\subsection{VAE and Vocoder}
We use a VAE model consisting of an encoder and decoder built with stacked convolutional modules \cite{kingma2022a}. The model is retrained on clean speech, noisy speech, and background noise data, following the loss functions described in \cite{liu2023a}. For the vocoder, we employ BigVGAN \cite{lee2023b} to generate speech samples from the enhanced mel-spectrogram. This model is retrained using only clean speech, according to the original work's settings. Both the VAE and BigVGAN models are trained independently, and their parameters are kept frozen during the training of cLDM.
\begin{table}
\center
\footnotesize
\caption{Model configuration for VAE and cLDM. }
\setlength\tabcolsep{3.0pt} 
\begin{tabular}{lcc}
\specialrule{.1em}{.05em}{.05em}
{Model} & {VAE} & {cLDM} \\
 \hline
Number of Parameters		     	&83M          &866M          \\      
In/Out Channels 					&1/1		  &16/8	          \\
Latent Channels 					&8			  &8             \\ 
Number of Down/Up Blocks 			&4/4		 &4/4      \\ 
Block Out Channels 					&(128, 256, 512, 512)		&(320, 640, 1280, 1280)            \\
Activate Function         			&SiLU		&SiLU            \\
Attention Heads        			    &Null		&8            \\
Cross Attention Dimension         &Null		&1024           \\
 \hline
\specialrule{.1em}{.05em}{.05em}
\end{tabular}
\vspace{-6mm}
\end{table}
\section{Experiment}
\subsection{Dataset}
Since generative models benefit from large-scale training data, we selected the LibriSpeech corpus \cite{Panayotov2015}, using the 'train-clean-360' subset for training clean speech and 'test-clean' for testing speech. The training set comprises 360 hours of speech. For noise data, we utilized the AudioSet corpus\cite{Gemmeke2017} and filtered out audio clips containing human speech to ensure 'clean' noise. We carefully selected noise types to maintain a balanced noise dataset, resulting in a total of 250 hours of noise. Five noise types—laughing, gunshot, singing, car engine, and rain—were reserved as unseen noises for testing. We generated 3,000 hours of noisy-clean pairs and 1,000 hours of noisy-noise pairs with 95\% for training and 5\% for development. Note that the noisy-noise pairs are only used for the dual-context learning framework. For the test set, we created a 2-hour dataset with seen noise and a 2-hour dataset with unseen noise, allocating 24 minutes to each noise type. During data generation, the SNR was randomly selected between -5 dB and 15 dB with a uniform distribution. To further demonstrate the generalization capability of our proposed model, we also include two mismatched test sets from VoiceBank+DEMAND \cite{Botinhao2016} and DNS Challenge 2020 \cite{Reddy2020} datasets. The VoiceBank+DEMAND test set consists of 824 noisy-clean pairs from 2 speakers. The DNS Challenge 2020 test set consists of 150 noisy-clean pairs.  All audio samples were resampled to 16 kHz.  
\subsection{Evaluation Metrics}
For performance evaluation, we use three intrusive measures: Perceptual Evaluation of Speech Quality (PESQ) \cite{Beerends2001}, extended short-term objective intelligibility (ESTOI) \cite{Jensen2016A}, and scale-invariant signal-to-distortion ratio (SI-SDR) \cite {Roux2019S} to assess speech quality, intelligibility, and noise removal, respectively. Additionally, we employ two non-intrusive measures: Wav2Vec MOS (WV-MOS) \cite{Andreev2023} and Deep Noise Suppression MOS (DNSMOS) \cite{reddy2022dn} to evaluate speech quality based on DNN models.
\subsection{Baselines}
We compare our proposed latent diffusion-based method with three closely related diffusion-based speech enhancement methods: CDiffuSE \footnote{CDiffuSE: https://github.com/neillu23/CDiffuSE}, SGMSE+  \footnote{SGMSE+: https://github.com/sp-uhh/sgmse}, and StoRM \footnote{StoRM: https://github.com/sp-uhh/storm}. CDiffuSE is a time-domain generative method with a conditional diffusion process, while both SGMSE+ and StoRM are score-based generative methods utilizing stochastic diffusion in the STFT domain. We adopted the open-sourced recipes  and retrained the models using our training set. Additionally, we evaluate our dual-context learning scheme against the NASE approach, which employs a pre-trained noise classifier (BEATs \cite{Chen2022B}) as a noise encoder to extract noisy acoustic embeddings as noise conditioners. Our proposed method is named cLDM+DCL, and we also create a new baseline, cLDM+NASE, by replacing DCL with NASE in our approach. Two discriminative methods, Conv-TasNet \cite{Luo2019} and MetricGAN+ \cite{Fu2022M}, are included as additional baselines.
\begin{table}
\center
\footnotesize
\caption{Ablation study for the effect of the number of reverse process steps $T$ and the DCL scheme}
\setlength\tabcolsep{3.0pt} 
\begin{tabular}{lccccccc}
\specialrule{.1em}{.05em}{.05em}
{Model}&{Steps (T)} & {PESQ} &{ESTOI} & {SI-SDR} & {WV-MOS} & {DNS-MOS} &{RTF}\\ \hline
Unprocessed	&-	       &1.28        &0.62      &3.40    &1.63    &2.88   &-      \\ \hline
&10		           	   &2.64        &0.87      &16.5    &3.68    &3.50   &\textbf{0.09}    \\ 
&20 				   &2.66		&0.87	   &16.6     &3.70     &3.51    &0.16   \\
cLDM+DCL&30 		   &2.68		&0.87      &16.8     &3.72    &3.53    &0.23    \\
&40 				  &2.70         &\textbf{0.88}      &17.0     &\textbf{3.74}    &\textbf{3.55}    &0.32   \\ 
&50 				  &\textbf{2.71}&\textbf{0.88}      &\textbf{17.1}     &3.73    &3.54      &0.40   \\
 \hline
cLDM&50		           &2.64        &0.86      &16.4    &3.69    &3.51   &0.40 \\ \hline
\specialrule{.1em}{.05em}{.05em}
\end{tabular}
\vspace{-6mm}
\end{table}   
\subsection{Traning Details}
The model configurations for the VAE and cLDM components used in our cLDM+DCL model are detailed in Table 1. Our VAE compresses a mel-spectrogram of dimensions  $1 \times L \times F$  into a latent representation of size $8 \times \frac{L}{4} \times \frac{F}{4}$ with $C=8$ and $r=4$. The VAE was trained on 360 hours of clean speech, 250 hours of noise, and 3,000 hours of generated noisy speech. Our training was conducted using the AdamW optimizer \cite{loshchilov2019d} with a learning rate of $4.5\times 10^{-6}$ and a batch size of 6, over 1 million steps on a single NVIDIA A800 GPU. Additionally, we trained BigVGAN as our vocoder using 360 hours of clean speech, with a window size of 1024 and a hop size of 160, according to the official recipe guidelines \footnote{https://github.com/NVIDIA/BigVGAN}. BigVGAN was also trained for 1 million steps on a single GPU. We extracted 64-band mel-spectrograms for VAE and BigVGAN. Our cLDM architecture is based on the Stable Diffusion U-Net \cite{Ho2020D}, consisting of 866 million parameters. We used 8 channels and set the cross-attention dimension to 1024 in the U-Net model. We used $T=1000$ steps in the forward process and $T=50$ steps in the reverse process for final evaluation. The AdamW optimizer was employed with a learning rate of $3e-5$ and a linear learning rate scheduler. The cLDM was trained on 10-second audio segments, with a batch size of 32, over 2 million steps on 2 NVIDIA A800 GPUs. Noisy-clean pairs and noisy-noise pairs were randomly selected during training. The text embeddings use a size of 768 from the T5 model. For all baseline models, we followed the training guidelines specified in their respective works. We chose the best models based on the validation results on the development set.  
\subsection{Ablation Study}
We begin by evaluating the impact of the number of reverse process steps, $T$, and the Dual-Context Learning (DCL) scheme. Table 2 compares model performance with $T = 10, 20, 30, 40, 50$, both with and without DCL. In the DCL case, our cLDM+DCL model is trained on both noisy-clean and noisy-noise pairs, while the model without DCL, referred to as cLDM, is trained solely on noisy-clean pairs. We observe that cLDM+DCL benefits from an increased number of reverse diffusion steps, maintaining strong performance even with as few as 10 steps. The real-time factor (RTF) remains low, starting at 0.09 and only reaching 0.4 at 50 steps. In contrast, cLDM without DCL at 50 steps performs similarly to cLDM+DCL at just 10 steps, highlighting DCL’s effectiveness in improving the model's discriminative capability.
\begin{table}
\center
\footnotesize
\caption{Performance comparison on seen-noise test set. }
\setlength\tabcolsep{3.0pt} 
\begin{tabular}{lccccccc}
\specialrule{.1em}{.05em}{.05em}
\multirow{2}{*}{Model} & \multirow{2}{*}{Type} &\multicolumn{5}{c}{Evaluation Metrics} &\multirow{2}{*}{RTF}\\ 
\cline{3-7} 
& & {PESQ} &{ESTOI} & {SI-SDR} & {WV-MOS} & {DNS-MOS} & \\ \hline
Unprocessed		  &-  &1.28      &0.62      &3.40    &1.63   &2.88  &-     \\ \hline
Conv-TasNet \cite{Luo2019} 	 &D &2.59		&0.85	    &\textbf{18.5}   &3.65   &3.45   &\textbf{0.15}   \\
MetricGAN+ \cite{Fu2022M} 		 &D &\textbf{2.85}		&0.83       &7.8    &3.45   &3.42   &0.35    \\ \hline
CDiffuSE \cite{Lu2022C} 		 &G &2.28		&0.79       &11.8   &3.02   &3.00   &0.89   \\ 
SGMSE+ \cite{Richter2023S} 			 &G &2.62		&0.87       &16.1   &3.64   &3.45   &2.15   \\
StoRM \cite{Lemercier2022}           &G &2.63		&0.87       &16.5   &3.66   &3.48   &1.98    \\ \hline
cLDM+NASE       &G &2.66		&\textbf{0.88}       &17.0   &3.69   &3.52   &0.87    \\
cLDM+DCL        &G &2.71		&\textbf{0.88}       &17.1   &\textbf{3.73}   &\textbf{3.54}   &0.40    \\
 \hline
\specialrule{.1em}{.05em}{.05em}
\end{tabular}
\vspace{-6mm}
\end{table}
\begin{table}
\center
\footnotesize
\caption{Performance comparison on unseen-noise test set. The PESQ metric was used for the performance evaluations}
\setlength\tabcolsep{3.0pt} 
\begin{tabular}{lccccccc}
\specialrule{.1em}{.05em}{.05em}
\multirow{2}{*}{Model} & \multirow{2}{*}{Type} &\multicolumn{5}{c}{Unseen Noise Types} &\multirow{2}{*}{RTF}\\ 
\cline{3-7} 
& & {laughing} &{gunshot} & {singing} & {car engine} & {rain} & \\ \hline
Unprocessed		 &- &1.35       &1.19     &1.05   &1.12     &1.37  &-   \\ \hline
Conv-TasNet \cite{Luo2019} 	&D &2.36	   &2.14	  &1.90   &2.18    &2.28   &\textbf{0.15}   \\
MetricGAN+ \cite{Fu2022M} 		&D &2.58	   &2.42     &2.30   &2.50    &2.70   &0.35    \\ \hline
CDiffuSE \cite{Lu2022C} 		&G &2.26	   &2.21     &2.03   &2.11    &2.30   & 0.89   \\ 
SGMSE+ \cite{Richter2023S}			&G &2.60	  &2.57     &2.43    &2.51    &2.61    &2.15   \\
StoRM \cite{Lemercier2022}          &G &2.62 	  &2.59       &2.45   &2.58    &2.63   &1.98    \\ \hline
cLDM+NASE      &G &2.65 	  &2.62       &2.50   &2.62    &2.67   &0.87    \\
cLDM+DCL       &G &\textbf{2.70} 	  &\textbf{2.65}       &\textbf{2.55}   &\textbf{2.68}    &\textbf{2.71}   &0.40    \\
 \hline
\specialrule{.1em}{.05em}{.05em}
\end{tabular}
\vspace{-6mm}
\end{table}
\subsection{Results and Discussion}
We then compare the performance of our proposed cLDM+DCL approach with baseline models across different test set configurations. Table 3 presents the performance comparison on the seen-noise test set. It is observed that the discriminative models excel on the metrics they were specifically optimized for, with Conv-TasNet achieving the best SI-SDR score and MetricGAN+ achieving the highest PESQ result. However, these models couldn't maintain consistent performance across multiple measures. In contrast, the generative methods exhibit more balanced performance across all five metrics. Except for CDiffuSE, all other generative methods produced competitive or superior VW-MOS and DNS-MOS results compared to the predictive methods. Generally, discriminative methods have lower RTFs than generative methods. In cLDM+DCL, the use of latent representations helps reduce the RTFs equivalent to that of MetricGAN+. 
Among the generative approaches, cLDM+NASE and cLDM+DCL outperform CDiffuSE as well as score-based methods of SGMSE+ and StoRM, demonstrating the effectiveness of applying diffusion generation process on low-dimensional latent spaces compared to high-dimensional spaces. Regarding the impact of noise exploration, cLDM+DCL outperforms cLDM+NASE, revealing the effectiveness of learning noise distributions for enhanced speech enhancement.

Next, we present in Table 4 the performance comparison results of our cLDM+DCL with various baselines on the unseen-noise test set, using the PESQ metric for clarity. We observe that the discriminative methods show a significant decline in performance on this metric due to the mismatch in noise domains. In contrast, the generative methods achieve results comparable to those on the seen-noise test set, highlighting their robustness against unseen noises. Among the generative approaches, our cLDM+DCL outperforms all baselines across the unseen noise types, benefiting from the noise characteristics learned through the DCL scheme.

Finally, we present performance comparisons on two widely-used benchmarks: the VoiceBank+DEMAND test set and the DNS Challenge 2020 non-blind test set without reverberation. All models were trained on our simulated LibriSpeech+AudioSet data. We observe that while the baselines deliver solid performance, there is some decline in metrics compared to their originally reported results, which were trained on matched datasets. Although Conv-TasNet achieved the highest SI-SDR scores, our cLDM+DCL model outperformed Conv-TasNet and the others on the remaining metrics, particularly on perceptual metrics such as PESQ, WV-MOS, and DNS-MOS. This suggests that our cLDM+DCL offers advantages in enhancing the clarity and intelligibility of speech, rather than solely preserving point-wise fidelity \footnote{The audio samples are available at https://github.com/alibabasglab/cLDM-DCL}.
\begin{table}
\center
\footnotesize
\caption{Performance comparison on out-of-domain VoiceBank+DEMAND test set. }
\setlength\tabcolsep{3.0pt} 
\begin{tabular}{lccccccc}
\specialrule{.1em}{.05em}{.05em}
\multirow{2}{*}{Model} & \multirow{2}{*}{Type} &\multicolumn{5}{c}{Evaluation Metrics} &\multirow{2}{*}{RTF}\\
\cline{3-7}
& & {PESQ} &{ESTOI} & {SI-SDR} & {WV-MOS} & {DNS-MOS} & \\ \hline
Unprocessed		  &-  &1.97      &0.79      &8.40    &2.99   &3.09   &-     \\ \hline
Conv-TasNet \cite{Luo2019} 	 &D  &2.52		&0.82	    &\textbf{18.2}   &4.16   &3.19   &\textbf{0.15}   \\
MetricGAN+ \cite{Fu2022M} 		 &D &2.82		&0.80       &8.00    &3.82   &3.20   &0.35    \\ \hline
CDiffuSE \cite{Lu2022C} 		 &G &2.38		&0.77       &12.2   &3.55   &2.95   &0.89   \\ 
SGMSE+ \cite{Richter2023S} 			 &G &2.85		&0.83       &16.5   &4.13   &3.49   &2.15   \\
StoRM \cite{Lemercier2022}           &G &2.85		&0.84       &17.1   &4.17   &3.50   &1.98    \\ \hline
cLDM+NASE       &G &2.87		&0.85       &17.5   &4.21   &3.53   &0.87    \\
cLDM+DCL        &G &\textbf{2.89}		&\textbf{0.86}       &17.8   &\textbf{4.25}   &\textbf{3.55}   &0.40    \\
 \hline
\specialrule{.1em}{.05em}{.05em}
\end{tabular}
\vspace{-6mm}
\end{table}
\begin{table}
\center
\footnotesize
\caption{Performance comparison on out-of-domain DNS Challenge 2020 non-blind test set. }
\setlength\tabcolsep{3.0pt} 
\begin{tabular}{lccccccc}
\specialrule{.1em}{.05em}{.05em}
\multirow{2}{*}{Model} & \multirow{2}{*}{Type} &\multicolumn{5}{c}{Evaluation Metrics} &\multirow{2}{*}{RTF}\\
\cline{3-7}
                 & & {PESQ} &{ESTOI} & {SI-SDR}            & {WV-MOS} & {DNS-MOS} & \\ \hline
Unprocessed		  &-  &1.58      &0.92     &9.07            &1.79       &3.16  &-     \\ \hline
Conv-TasNet \cite{Luo2019} 	 &D &2.56		&0.93	&\textbf{18.9}       &2.51        &3.55   &\textbf{0.15}   \\
MetricGAN+ \cite{Fu2022M} 		 &D &2.86		&0.95       &8.60           &2.17        &3.62   	  &0.35    \\ \hline
CDiffuSE \cite{Lu2022C} 		 &G &2.43		&0.93       &13.5           &2.38        &3.37       &0.89   \\ 
SGMSE+ \cite{Richter2023S} 			 &G &2.88		&0.96       &17.4           &3.01        &3.77       &2.15   \\
StoRM \cite{Lemercier2022}           &G &2.89		&0.96       &17.6           &3.07        &3.79       &1.98    \\ \hline
cLDM+NASE       &G &2.91		&0.96       &18.2           &3.23        &3.82       &0.87    \\
cLDM+DCL        &G &\textbf{2.95}	&\textbf{0.97}  &18.5 &\textbf{3.29}   &\textbf{3.85}   &0.40    \\
 \hline
\specialrule{.1em}{.05em}{.05em}
\end{tabular}
\vspace{-6mm}
\end{table} 
\section{Conclusions}
We introduced a generative framework, cLDM+DCL, by integrating a conditional latent diffusion model with dual-context learning for speech enhancement. The proposed cLDM operates in a low-dimensional latent space, reducing complexity and improving the efficiency of the generation process. The DCL scheme further strengthens the model’s ability to handle diverse and unseen noise environments. Experimental results on large dataset validated the effectiveness of our proposed approach.
\bibliographystyle{IEEEbib}
\bibliography{refs}

\end{document}